\documentclass[reprint,floatfix,
showpacs,preprintnumbers,
nofootinbib,
nobibnotes,
amsmath,amssymb,aps,prd
]{revtex4-2}

\usepackage{multirow,hyperref}%
\usepackage[dvipsnames]{xcolor}%
\usepackage{multirow}%
\usepackage[final]{microtype}%
\usepackage{subcaption}%
\usepackage{amsmath,amssymb,amsfonts}%
\usepackage{amsthm}%
\usepackage{mathrsfs}%
\usepackage[title]{appendix}%
\usepackage{xcolor}%
\usepackage{textcomp}%
\usepackage{booktabs}%
\usepackage{algorithm}%
\usepackage{algorithmicx}%
\usepackage{algpseudocode}%
\usepackage{listings}%
\usepackage{graphicx}
\usepackage{dcolumn}
\usepackage{bm}

\def\be{\begin{equation}}
\def\ee{\end{equation}}
\def\beq{\begin{eqnarray}}
\def\eeq{\end{eqnarray}}
\def\beqs{\begin{subequations}}
\def\eeqs{\end{subequations}}

\def\na{\nabla_a}
\def\nb{\nabla_b}
\def\nc{\nabla_c}
\def\Ld{\pounds_{\vec{\phi}}}
\def\bg{\bar{g}}
\def\bL{\bar{\Lambda}}
\def\SS{\mathsf{S}}
\def\R{\mathcal{R}}

\def\sk{\smallskip \noindent}
\def\n{\nonumber}

\begin{document}




\title{Lovelock type brane gravity from a minimal surface perspective}


\author{Efra\'\i n Rojas}
\email{efrojas@uv.mx}

\author{G. Cruz}
\email{giocruz@uv.mx}

\affiliation{Facultad de F\'\i sica, Universidad Veracruzana, 
Paseo No. 112, Desarrollo Habitacional Nuevo Xalapa, Xalapa-Enr\'\i quez, 91097, Veracruz, M\'exico
}%




\date{\today}


\begin{abstract}
We explore the correspondence between the parallel 
surfaces framework, and the minimal surfaces framework, 
to uncover and apply new aspects of the geometrical and mechanical content behind the so-called Lovelock-type 
brane gravity (LBG). We show how this type of brane 
gravity emerges naturally from a Dirac-Nambu-Goto (DNG) 
action functional built up from the volume element 
associated with a world volume shifted a distance $\alpha$ 
along the normal vector of a germinal world volume, and 
provide all known geometric structures for such a theory. 
Our development highlights the dependence of the geometry 
for the displaced world volume on the fundamental forms, 
as well as on certain conserved tensors, defined on the 
outset world volume. Based on this, LBG represents a 
natural and elegant generalization of the DNG theory to 
higher dimensions. Moreover, our development 
allows for exploring disformal transformations in Lovelock 
brane gravity and analyzing their relations with scalar-tensor 
theories defined on the brane trajectory. Likewise, this geometrical correspondence would enable us to establish 
contact with tractable Hamiltonian approximations for this 
brane gravity theory, by exploiting the linkage with a DNG model, and thus start building a suitable quantum version.

\end{abstract}


\maketitle


\section{Introduction}
\label{sec:intro}

Branes are natural bricks in many higher-dimensional field theories. 
On the technical level these are extended objects of any dimension 
which are generalizations of particles and strings which attempt 
to represent many physical systems of an appropriate dimension, in 
terms of fields confined to their trajectories (world volume), 
propagating in a fixed background~\cite{Carter1992,defo1995}.  
In a geodetic setting, with no matter included, a brane can be slack 
and may wiggle and move, but its world volume will take on a 
certain shape, so the only relevant degrees of freedom (dof) should 
be those associated with its geometric configuration depending on 
how the world volume is embedded within the ambient spacetime. On the
dynamical level, this fact leads to analyze its behaviour through 
Lagrangians constructed locally from the geometry of the world volume 
through geometrical invariants, composed from the fundamental 
forms associated to this surface, $L (g_{ab}, K_{ab})$. The presence 
of the extrinsic curvature signals the existence of second-order
temporal and spatial derivatives of the field variables, the 
embedding functions $X^\mu$. In a relativistic context this produces 
reluctance due to the emergence of non-physical dof that arise 
as a result of handling usual fourth-order equations of motion 
(eom) and therefore dealing with an unexpected number of dof. 
One may wonder, what conditions in such Lagrangians must be 
fulfilled to ensure that eom do not contain derivatives of $X^\mu$ 
higher than second-order? 
The Regge-Teitelboim (RT) brane model falls into this category of 
gravity theories, and plays a key role in the understanding of 
richer geometric models leading to second-order equations of motion~\cite{RT1975}. Guided by Lanczos-Lovelock 
gravity~\cite{Lovelock1971}, and accompanied by the 
Gauss-Codazzi and Codazzi-Mainardi integrability conditions for surfaces~\cite{defo1995,Spivak1970}, 
a tempting alternative picks out only those terms composed 
of appropriate antisymmetric products of the fundamental forms~\cite{Rojas2013,Rojas2016,Rojas2019}.

Lovelock type brane gravity describes extended objects 
moving geodetically in a higher-dimensional flat spacetime, 
characterized by second-order geometrical scalars, and retaining 
second-order field equations. 
The antisymmetric products indicated in~(\ref{lbi})
, similar in form to the original Lovelock theory, involve the 
first and second fundamental forms so that, within this framework 
of extended objects, a larger number of geometric scalars appear, 
with the interesting feature of providing a second-order 
equation of motion. This aspect makes the theory free 
from many of the pathologies that plague higher-order derivative 
theories thus ensuring no propagation of extra dof. The price 
to pay is that the resulting equation of motion is highly 
non-linear in the field variables $X^\mu$. The theory has led 
to interest in having potential physical applications, mainly at the cosmological level, since it allows for  considering alternative purely geometrical theories that might underlie the current puzzle of cosmic acceleration~\cite{Rojas2012,Rojas2024}.

There is only a field equation that results in an extension of 
the original Lovelock tensor~\cite{Lovelock1971}, but in our 
scheme, contracted with the extrinsic curvature tensor of the 
embedding, $\sum_n J^{ab}_{(n)} K^{ab} = 0$, representing thus 
a generalization of the original Lovelock equations in the sense 
that the Lovelock brane equation is fulfilled for every 
solution of the pioneer Lovelock equations. 
Based on this, the theory has a particular built-in Lovelock limit.
The Lovelock type brane scalars are similar in form either 
to the original form of the Lovelock invariants in gravity or 
to their counterterms necessary in order to have a well-posed 
variational problem~\cite{Lovelock1971,Myers1987,Davis2003,Olea2005,Padmanabhan2017}. 
In this parlance, we must proceed with caution to avoid a 
misimpression of the theory. For even values of these invariants, 
they resemble the usual Gauss-Bonnet invariants while for odd 
values, the corresponding Lovelock brane invariants resemble 
the Gibbons-Hawking-York-Myers boundary terms which are seen 
as counterterms if we have the presence of bulk Lanczos-Lovelock 
invariants. As discussed in~\cite{Rojas2016,Rojas2019}, the odd terms 
involve both time and spatial derivatives of the field 
variables thus contributing to the brane dynamics contrary 
to what occurs in the pure Lanczos-Lovelock gravity theory 
where the counterterms contain only spatial derivatives 
of the metric components.  To accommodate such Lovelock type 
Lagrangians it is evident that more dimensions than four are 
needed. Within this assumption, one must not overlook the local 
isometric embedding theorem~\cite{Friedman1961,Rosen1965}. 

It is interesting and inquiring that the equation of motion 
of LBG resembles the minimality condition for surfaces so that a 
natural question in this direction is whether there exists 
a particular immersion leading to a DNG type setup with the 
outlined equation of motion. In this sense, the parallel 
surfaces framework provides the geometric 
engineering to answer the question~\cite{Eisenhart1940,Wintner1952}.
On the basis of minimal surfaces development, the variation 
of a DNG action functional yields that the mean extrinsic 
curvature, $K^*$, of a manifold $m^*$ parallel to a given 
one $m$, vanishes identically. According to our assumption, 
this condition would be traded for a series involving the 
contraction of handy conserved brane tensors with the 
extrinsic curvature defined on the primordial world volume $m$.

In this paper we show how the parallel surfaces framework~\cite{Eisenhart1940,Wintner1952}, adapted to the 
scheme of extended objects of arbitrary dimensions, gives 
rise to the named Lovelock type brane gravity theory. 
Backed by matrix techniques we prove that the Lovelock 
type brane invariants (LBI) form the elements of a finite 
series expansion of principal minors of $\det (K^a{}_b)$ 
relative to a matrix transformation relating the tangent 
basis frame of a parallel world volume $m^*$, obtained by 
laying off equal distance $\alpha$ along a normal vector $n^\mu$ 
being the unit normal to a world volume $m$, and the tangent 
basis frame of $m$. The main geometrical structures defined 
on $m^*$, as seen from $m$, are expressed in terms of the 
fundamental forms of $m$ and of some conserved tensors 
$J^{ab}_{(n)}$ that form the backbone of the linear 
momentum density of the extended object. We believe that 
our approach could lead to a more tractable Hamiltonian 
formalism for the LBG, where the RT model is 
included~\cite{RT1975}, and thus pave the way to establish 
contact with known quantum approximations relative to a DNG action~\cite{Neeman1995,Ansoldi2000,Pavsic2016}. A survey of 
literature shows that a plenty of mathematical works exists
on the parallel surfaces approach, but as far as we know, 
at least in the relativistic extended objects framework, its 
impact is not completely clear. In this sense, 
this paper intends to bridge a number of gaps concerning the 
correspondence between parallel surfaces scheme and Lovelock 
type brane gravity. Higher-codimensions development will not 
be not considered here and is beyond the scope of this work. 

This paper is organized as follows. In Sec.~\ref{sec2} we 
quickly recap the geometrical structures defined on the brane 
trajectory and their variational properties to understand its 
mechanical behavior. In Sec.~\ref{sec3} we accommodate the 
parallel surfaces framework and develop its geometry in 
the extended objects setting. Further, we introduce 
a DNG type action written in terms of the geometry of a parallel 
world volume $m^*$ and recall the results associated with 
its dynamical evolution. Sec.~\ref{sec4} is devoted to 
analyzing the close relationship between a DNG theory and 
LBG, and the existence of disformal transformations when 
considering a varying distance for parallel surfaces.
In Sec.~\ref{sec5} we include arbitrary matter fields 
confined to the world volume $m$ and conjecture about the 
appearance of fictional matter encoded as dark matter. 
Finally, in Sec.~\ref{sec6} we conclude with some remarks 
for further work.

\section{Extended objects framework}
\label{sec2}

Consider a $p$-dimensional space-like brane, $\Sigma$, floating
in a flat Minkowski spacetime, $\mathcal{M}$, of dimension $N=p+2$
with metric $\eta_{\mu\nu}$, \,\,($\mu,\nu = 0,1,2,\ldots, p+1$). The trajectory of $\Sigma$ (world volume), denoted by $m$, is the 
focus of attention as a sufficiently smooth $(p+1)$-dimensional 
time-like manifold. This is an oriented hypersurface manifold 
described by the $N$ embedding functions of the world volume 
local coordinates $x^a$
\be 
X^\mu(x^a), \qquad \quad a= 0,1,2,\ldots, p.
\label{fembedding}
\ee
The tangent space to $m$ is spanned by the $p+1$ tangent 
vectors $e^\mu{}_a := \partial_a X^\mu$ while the normal 
space is one-dimensional and is spanned by a single space-like 
vector $n^\mu$. This orthonormal basis is defined through 
the relations $e_a \cdot n = \eta_{\mu\nu} e^\mu{}_a n^\nu 
= 0$ and $n\cdot n = \eta_{\mu\nu} n^\mu n^\nu = 
1$. Hereafter, a central dot will denote contraction with the 
Minkowski metric.  
 
The basis $\{ e_a, n \}$ induces the first and the second 
fundamental forms 
\be 
\begin{aligned}
g_{ab} &= \partial_a X \cdot \partial_b X,
\\
K_{ab} &= \partial_a X \cdot \partial_b n,
\end{aligned}
\label{FSff}
\ee
commonly known as the induced metric and the extrinsic
curvature, respectively, and the not least important, but 
frequently overlooked, third fundamental form
\be 
\SS_{ab} = \partial_a n \cdot \partial_b n.
\label{Tff}
\ee
Fundamental forms characterize any surface whereby all 
geometrical world volume invariants can be generated from 
$g_{ab}$, $K_{ab}$ and $\SS_{ab}$.

Using the Gauss-Weingarten (GW) equations, namely
$\partial_a e_b = \Gamma^c_{ab} e_c - K_{ab} \,n$ 
and $\partial_a n = K_{ab} g^{bc} e_c$,~\cite{Spivak1970}, 
where $\Gamma^c_{ab}$ denotes the connection coefficients 
on $m$, one observes that $\SS_{ab}$ is expressible 
entirely in terms of the first and second fundamental 
forms, that is, $\SS_{ab} = K_{ac}g^{cd} 
K_{bd}$. 
From this fact, such structure along with its variation,
helps to write many of the expressions used in this work
in a more compact way. On physical grounds, its trace
$\SS = g^{ab} \partial_a n \cdot \partial_b n$ together 
with the constraint $n\cdot n =1$, can be considered
as a non-linear sigma model living on the curved geometry
of the world volume~\cite{Capo2005}. The connection 
coefficients are calculated from 
$\Gamma^c_{ab} = g^{cd} \left( e_d \cdot \partial_a 
e_b \right)$. Further, by $\na$ we will denote the covariant 
derivative compatible with the induced metric, $\na g_{bc} = 0$.
 
The infinitesimal changes of $m$ through $X^\mu (x^a)
\rightarrow X^\mu (x^a) + \delta X^\mu (x^a)$, can be decomposed
into tangential and normal deformations as $\delta X^\mu = \phi^a
(x^b) e^\mu{}_b + \phi (x^b) n^\mu$, where $\phi^a$ and $\phi$ 
denote tangential and normal deformation fields, 
respectively. The variations we shall need 
are~\cite{defo1995}
\beq 
\delta e^\mu {}_a &=& \left( K_{ab}g^{bc} \phi + \na \phi^c 
\right) e_c + \left( - K_{ab} \phi^b + \na \phi \right) n^\mu
\label{var0}
\\
\delta n^\mu &=& \left( K_{ab} g^{ac}\, \phi^b  - g^{ac} \na \phi
\right) e^\mu{}_c, 
\label{var1}
\eeq
for the world volume basis while 
\beq 
\delta g_{ab} &=& 2 K_{ab} \,\phi + \Ld \,g_{ab},
\label{var1a}
\\
\delta K_{ab} &=& - \na \nb \phi + \SS_{ab}\,\phi + \Ld K_{ab},
\label{var1b}
\\
\delta \SS_{ab} &=& - 2 K_{(a}{}^c \nabla_{b)} \nc \phi +
\Ld \SS_{ab},
\label{var1c}
\eeq
for the fundamental forms where $\Ld$ stands for the Lie 
derivative along the vector field $\phi^a$. This acts on 
the fundamental forms as follows
\be 
\begin{aligned}
\Ld g_{ab} &= \phi^c \nc g_{bc} + 2 g_{c(a} \nabla_{b)} \phi^c,
\\
\Ld K_{ab} &= \phi^c \nc K_{ab} + 2K_{c(a} \nabla_{b)} \phi^c,
\\
\Ld \SS_{ab} &= \phi^c \nc \SS_{ab} + 2 \SS_{c(a} \nabla_{b)} 
\phi^c.
\end{aligned}
\label{Ldffs1}
\ee
In connection with~(\ref{var1a}-\ref{var1c}), further useful 
relationships are given by
\beq
\delta g^{ab} &=& - 2 K^{ab} \,\phi + \Ld g^{ab},
\label{var1d}
\\
\delta K^{ab} &=& - \nabla^a \nabla^b \phi - 3 \mathsf{S}^{ab}\,\phi 
+ \Ld K^{ab},
\label{var1e}
\\
\delta \SS^{ab} &=& - 2 K^{(a}{}_c \nabla^{b)} \nabla^c \phi 
- 4 K^{(a}{}_c \SS^{b)c} \,\phi + \Ld \SS^{ab}.
\label{var1f}
\eeq
It should be stressed that the tangential deformations can 
always be associated with infinitesimal reparametrizations, and 
can be ignored safely if there is no boundary.

\section{Parallel surfaces framework for extended objects}
\label{sec3}

The parallel surfaces framework~\cite{Eisenhart1940,Wintner1952} 
appears to provide a purely geometrical support for LBG. To 
prove this let us start by assuming a manifold $m^*$ determined by the following embedding functions
\be 
X^{*\mu} (x^a) = X^\mu (x^a) + \alpha\,n^\mu (x^a),
\label{sembedding}
\ee
where $\alpha$ is a constant, $X^\mu$ is given by (\ref{fembedding}), 
and $n^\mu$ is the unit normal vector to $m$. Manifold $m^*$ 
represents a geometrically parallel world volume to $m$. Here, 
and henceforth, starred quantities will denote geometric
structures defined on $m^*$.

It is worth noting that if $m$ and $m^*$ are parallel 
manifolds, separated by equal distances $\alpha$ along 
the normal $n^\mu$, a family of geometrically parallel 
world volumes can be produced by varying the parameter 
$\alpha$ in~(\ref{sembedding}). In this spirit, from an 
opposite point of view, $m$ is also parallel to $m^*$ 
taking $m^*$ itself as the origin. A word of caution is 
needed. It is clear that there is an endless number of 
$m^*$s parallel to a given $m$, but to maintain the 
right causal structure for $m^*$, it is mandatory to 
assume appropriate values of $\alpha$.

The tangent space of $m^*$ is spanned by the vectors 
$E^\mu{}_a := \partial_a X^{*\mu}$, that is
\be 
E^\mu{}_a = e^\mu{}_a + \alpha K_{ab}g^{bc} e^\mu{}_c,
\label{tang1}
\ee
where we have used the GW equations already outlined. 
It follows that $e^\mu{}_a$ can be taken as a basis of 
the tangent space at $X^{*\mu}$. It can be readily 
proved that the unit normal vector $n^{*\mu}$ to $m^*$ 
coincides with $n^\mu$, the unit normal to $m$. The 
new orthonormal basis, $\{ E^\mu{}_a, n^\mu \}$ is defined 
through the identities $E_a \cdot n = 0$ and $n \cdot 
n = 1$. 

In analogy to the induced metric on $m$,~(\ref{FSff}), 
considering~(\ref{tang1}) the metric coefficients
 $g^*_{ab}$ of $m^*$ turn out to be
\be 
g^*_{ab} = E_a \cdot E_b = g_{ab} + 2\alpha \, K_{ab} +
\alpha^2 \,\SS_{ab},
\label{Fff2}
\ee
where $\SS_{ab}$ is given by~(\ref{Tff}). For this basis,
in view of~(\ref{sembedding}), the corresponding 
Gauss-Weingarten equations take the form
\beq 
\partial_a E^\mu{}_b &=& \Gamma^{*c}_{ab} \,E^\mu{}_c -
K^*_{ab} \,n^\mu,
\label{GW2a}
\\
\partial_a n^\mu &=& K^*_{ac} \bg^{*cb} \,E^\mu{}_b,
\label{GW2b}
\eeq
where, $\bg^{*ab}$ should be understood as the inverse of the 
metric $g^*_{ab}$ such that $\bg^{*ac} g^*_{cb}
= \delta^a{}_b$, and $\Gamma^{*c}_{ab}$ are the connection
coefficients compatible with the starred metric $g^*_{ab}$.
Additionally, $K^*_{ab}$ denotes the extrinsic curvature
of $m^*$ defined as $K^*_{ab} := - n \cdot \partial_a E_b$.

The extrinsic geometry of $m^*$ is determined as follows. According~(\ref{tang1}), and the GW equation~(\ref{GW2b}), 
the starred extrinsic curvature assumes the form 
\be 
K^*_{ab} = E_a \cdot \partial_b n = K_{ab} + \alpha 
\,\SS_{ab}.
\label{Sff2} 
\ee
In the same parlance, the coefficients $\SS^*_{ab}$ of 
the third fundamental form associated to $m^*$, do not suffer 
any change
\be 
\SS^*_{ab} = \partial_a n^* \cdot \partial_b n^* 
= \mathsf{S}_{ab}.
\label{Tff2}
\ee
It is noteworthy that the underlying geometry on $m^*$ 
given by~(\ref{tang1}),~(\ref{Fff2}),~(\ref{Sff2}),
and~(\ref{Tff2}), at a given point $x^a$ on $m^*$,
can be expressed entirely in terms of the values of the 
fundamental forms at the corresponding point $x^a$ on $m$. 

We can further derive an expression for the starred 
Christoffel symbol. From~(\ref{GW2a}), $\Gamma^{*c}_{ab} 
= \bg^{*cd} \left( E_d \cdot D_a E_b \right)$, and the 
expression defining $\Gamma^c_{ab}$, as well as~(\ref{FSff}) 
and~(\ref{tang1}), a detailed yet straightforward 
computation leads
\be 
\Gamma^{*c}_{ab} = \Gamma^c_{ab} + 2\alpha\, 
_{\tiny K}\Gamma^c_{ab} + \alpha^2 \, 
_{\SS}\Gamma^c_{ab},
\label{Gamma2}
\ee
where
\beq
_{\tiny K}\Gamma^c_{ab} &:=& \frac{1}{2} \bg^{*cd}
\left( \na K_{bd} + \nb K_{ad} - \nabla_d K_{ab}\right),
\\
_{\tiny \SS}\Gamma^c_{ab} &:=& \frac{1}{2} \bg^{*cd}
\left( \na \SS_{bd} + \nb \SS_{ad} - \nabla_d \SS_{ab}\right).
\eeq

The intrinsic and extrinsic geometries of $m^*$ must
satisfy consistency conditions. The starred Gauss-Codazzi and 
Codazzi-Mainardi integrability conditions, 
\beq 
\R^*_{abcd} &=& K^*_{ac} K^*_{bd} - K^*_{ad} K^*_{cb},
\label{starR}
\\
0 &=& \nabla^*_a K^*_{bc} - \nabla^*_b K^*_{ac},
\eeq
are obtained from~(\ref{GW2a}) and~(\ref{GW2b}), where 
$\R^*_{abcd}$ and $\R_{abcd}$ denote the Riemann tensor of 
$m^*$ and $m$, respectively, while $\nabla^*_a$ is the 
covariant derivative compatible with $g^*_{ab}$.

For clever handling of the variational properties of 
the starred geometrical structures, one key factor 
is to observe from~(\ref{tang1}) the existence of a 
linear transformation from the $m$ tangent basis to the
starred one. It induces a matrix representation for
the geometry of $m^*$ as follows
\beq 
E^\mu{}_a &=& \Lambda^b{}_a e^\mu{}_b,
\label{Emua}
\\
g^*_{ab} &=& \Lambda^c{}_a \Lambda^d{}_b g_{cd},
\\
K^*_{ab} &=& \Lambda^c{}_a K_{cb},
\label{Kabstar}
\\
\SS^*_{ab} &=& \SS_{ab},
\label{Sabstar}
\eeq 
provided by the transformation matrix defined as
\be 
\Lambda^a{}_b := \delta^a{}_b + \alpha K^a{}_b.
\label{matrixL}
\ee
This matrix is non-singular and written in terms
of the first and second fundamental forms associated
with $m$. The latter equations will prove very useful
to directly calculate and describe the mechanical and 
geometrical content of our approach.

Accordingly, if $\bL^a{}_b$ denotes the inverse matrix 
of $\Lambda^a{}_b$, such that $\bL^a{}_c \Lambda^c{}_b 
= \delta^a{}_b$, then it immediately follows that $e^\mu{}_a 
= \bL^b{}_a E^\mu{}_b$ together with the handy identities
\be 
\begin{array}{lll}
g^*_{ab} = \Lambda^c{}_a \Lambda^d{}_b g_{cd},
& \qquad \bg^{*ab} &= \bL^a{}_c \bL^b{}_d g^{cd},
\\
g_{ab} = \bL^c{}_a \bL^d{}_b g^*_{cd},
& \qquad g^{ab} &= \Lambda^a{}_c \Lambda^b{}_d \bg^{*cd}.
\end{array}
\label{ids1}
\ee

\sk
As to the determinant of the metric $g^*_{ab}$, we 
shall compute this in terms of the matrix~(\ref{matrixL}). 
Indeed, from the elementary identity 
$\det (AB) = \det (A) \det (B)$, we readily obtain 
$g^* := \det (g^*_{ab}) = g \,\Lambda^2$ where $\Lambda 
:= \det (\Lambda^a{}_b)$, so that $\sqrt{-g^*} = \sqrt{-g} 
\,\Lambda$.

For the role it will play in what follows, it is convenient
to obtain an expression for $\Gamma^{*c}_{ab}$ in terms
of~(\ref{matrixL}). Certainly,
\beq
\Gamma^{*c}_{ab} &=& \bg^{*cd} E_d \cdot \partial_a (\Lambda^e{}_b 
\,e_e),
\n
\\
&=&  \bg^{*cd} \Lambda^h{}_d \Lambda^e{}_b g_{hf} 
\Gamma^f_{ae} +  \bg^{*cd} \Lambda^e{}_d g_{ef}\, \partial_a
\Lambda^f{}_b.
\n
\eeq
If this is combined with relations~(\ref{ids1}), the
result is a helpful identity
\be 
\label{Gamma2a}
\Gamma^{*c}_{ab} = \bL^c{}_d \Lambda^e{}_b \Gamma^d_{ae}
+ \bL^c{}_d \, \partial_a \Lambda^d{}_b. 
\ee
In a like manner, multiplying the latter by $\Lambda^r{}_c$ 
followed by multiplication by $\bL^b{}_s$, and
relabeling the dummy indices, we find
\be 
\label{Gamma2b}
\Gamma^c_{ab} = \Lambda^c{}_d \bL^e{}_b \Gamma^{*d}_{ae}
+ \Lambda^c{}_d \partial_a \bL^d{}_b.
\ee

In addition to this, equipped with~(\ref{matrixL}),
expression~(\ref{starR}) yields
\beq 
\R^*_{abcd} &=& \Lambda^e{}_a \Lambda^f{}_b \R_{efcd},
\label{starR2}
\\
&=& \R_{abcd} + 2 \alpha K^e_{[a} \R_{|e|b]cd} + \alpha^2
K^e{}_a K^f{}_b \R_{efcd}.
\n
\eeq

The variation of the starred first fundamental 
form~(\ref{Fff2}), with the aid of 
expressions~(\ref{var1a}-\ref{var1c}), and~(\ref{Ldffs1}), 
leads to
\beq
\delta g^*_{ab} &=& \delta \left( g_{ab} + 2\alpha\,K_{ab}
+ \alpha^2\,\SS_{ab} \right),
\n
\\
&=& 2K_{ab} \, \phi - 2\alpha \na \nb \phi + 2\alpha \,
\SS_{ab} - 2\alpha^2 K_{(a}{}^c \nabla_{b)}\nc \phi 
\n
\\
&+& \Ld g^*_{ab}.
\n
\eeq
Equipped with matrix~(\ref{matrixL}), it follows, 
considering the definition for the starred second 
fundamental form~(\ref{Sff2}), that
\be 
\delta g^*_{ab} = 2 K^*_{ab}\,\phi -2\alpha\,
\Lambda^c{}_{(a} \nabla_{b)} \nc \phi + \Ld g^*_{ab}.
\label{var2a}
\ee

With regards to the variation of the starred second 
fundamental form~(\ref{Sff2}), we get
\beq 
\delta K^*_{ab} &=& \delta \left( K_{ab} + \alpha \,
\SS_{ab} \right),
\n
\\
&=& - \na \nb \phi - 2\alpha K_{(a}{}^c \nabla_{b)} \nc \phi
+ \SS_{ab} \,\phi + \Ld K^*_{ab}.
\n
\eeq
In terms of~(\ref{matrixL}), we deduce the expression
\be 
\label{var2b}
\delta K^*_{ab} = \na \nb \phi - 2 \Lambda^c{}_{(a} \nabla_{b)}
\nc \phi + \SS_{ab} \,\phi + \Ld K^*_{ab}.
\ee
Finally, since $n^{*\mu} = n^\mu$, there will 
be no change in the variation of the starred third fundamental 
form, that is, $\delta \SS^*_{ab} = \delta \SS_{ab}$.
Clearly, by turning off the distance $\alpha$, the 
variations~(\ref{var2a}) and~(\ref{var2b}) are reduced 
to~(\ref{var1a}) and~(\ref{var1b}), as expected.

To end this section, for the sake of completeness, 
we provide the variation of the matrix~(\ref{matrixL}). 
Indeed, by considering~(\ref{var1b}) one directly verifies 
that
\beq 
\delta \Lambda^a{}_b &=& - \alpha \left( g^{ac} \nc \nb \phi
+ \SS^a{}_b\,\phi \right) + \Ld \Lambda^a{}_b.
\eeq

\subsection{Starred Dirac-Nambu-Goto action}
\label{subsec3a}

We will now consider extended objects within the framework 
of parallel surfaces, subject to a DNG dynamics. Consider 
the local action
\be 
S[X^{*\mu}] = - \mu \int_{m^*} d^{p+1}x \, \sqrt{- g^*},
\label{action0}
\ee
where $\mu$ is a constant representing a $p$-tension of 
the extended object $\Sigma$, and $g^* := \det (g^*_{ab})$. 
On a technical side, this action is proportional to the 
volume of $m^*$. It follows, as an immediate consequence, 
from the variation of this action with respect to the 
field variables $X^{*\mu}$, the oriented and compact Euler-Lagrange equation~\cite{Carter1992,defo1995}
\be 
K^* := \bg^{*\,ab} K^*_{ab} = 0,
\label{eom0}
\ee 
which means that the stationary state represents, with a 
slight abuse of language, a minimal hypersurface $m^*$. 
Within this framework, and guided by the approach developed 
in~\cite{Capo2000}, there is an associated conserved stress tensor
\be 
f^{*\,a \mu} = - \mu \sqrt{-g^*} \bg^{*\,ab} \,E^\mu{}_b,
\label{f-0}
\ee
which is purely tangential to $m^*$. On physical grounds, 
due to the Poincar\'e symmetry of the background spacetime, 
this represents the linear momentum density of the brane
mediated by the world volume geometry. 
The conservation property of~(\ref{f-0}) leads to another 
strategy for obtaining the equation of motion. Indeed, 
as explained in detail in~\cite{Capo2000}, $n\cdot \na^* 
f^{*\,a} = 0$ yields~(\ref{eom0}).

In another light, if the action~(\ref{action0}) is viewed 
from the framework of the original world volume $m$, 
$S[X^\mu]$, it can be straightforwardly shown that the 
stationary state is provided by condition (\ref{eom0}). This is 
explicitly proved in Appendix~\ref{app1}.
Clearly,~(\ref{eom0}) represents a second-order differential 
equation in derivatives of $X^{*\,\mu}$. On the other hand, 
on account of GW~(\ref{GW2a}), from $-\sqrt{-g^*} \bg^{*ab} 
n \cdot \nabla^*_a E_b = 0$, we can express~(\ref{eom0}) in 
terms of $N$ second-order hyperbolic partial differential 
equations for $X^{*\,\mu}$, that is, as a set of conserved 
currents
\be
\partial_a \left( \sqrt{-g^*} \bg^{*ab} 
\partial_b X^{*\mu} \right) = 0.
\label{eom2}
\ee
representing the usual harmonicity condition for DNG
extended objects.

\section{Parallel surfaces framework germinates Lovelock brane gravity}
\label{sec4}

The volume element in~(\ref{action0}) may be written 
in terms of minors $K_{(s)}$ related to the transformation 
matrix $\Lambda^a{}_b$. To prove this, according to the 
fact that $g^* = g \,\Lambda^2$ as outlined immediately below 
of~(\ref{ids1}), on considering the usual formula for a 
characteristic determinant (see Appendix~\ref{app2}),
the DNG action~(\ref{action0}) gets expressed as
\be
S[X^\mu] =  - \mu \int_{m^*} \sqrt{-g} \,\left(  1 +
\sum_{s=1}^{p+1} \frac{\alpha^s}{s!} L_{s}(g_{ab},K_{ab})\right),
\label{action1}
\ee
where $L_s (g_{ab},K_{ab}) :=s! K_{(s)}$ with $K_{(s)}$ 
being the principal minors of $\det (K^a{}_b)$. 
Here and henceforth we shall absorb the differential 
$d^{p+1}x$ in the integral sign for short the notation.
The issue of computing $K_{(s)}$ is conveniently tackled
by applying generalized Kronecker delta (gKd)
techniques. The gKd is an alternating tensor defined as
\be 
\delta^{a_1 a_2 a_3 \cdots a_n}_{b_1 b_2 b_3 \cdots b_n}
=
\left|
\begin{matrix}
\delta^{a_1}{}_{b_1} & \delta^{a_1}{}_{b_2}
& \delta^{a_1}{}_{b_3} & \cdots &
\delta^{a_1}{}_{b_n}
\cr
\delta^{a_2}{}_{b_1} & \delta^{a_2}{}_{b_2}
& \delta^{a_2}{}_{b_3} & \cdots &
\delta^{a_2}{}_{b_n}
\cr
\vdots & \vdots & \ddots & \vdots & \vdots
\cr
\delta^{a_{n-1}}{}_{b_1} & \delta^{a_{n-1}}{}_{b_2}
& \delta^{a_{n-1}}{}_{b_3} & \cdots &
\delta^{a_{n-1}}{}_{b_n}
\cr
\delta^{a_{n}}{}_{b_1} & \delta^{a_{n}}{}_{b_2}
& \delta^{a_{n}}{}_{b_3} & \cdots &
\delta^{a_{n}}{}_{b_n}
\end{matrix}
\right|.
\ee
This allows us to express
\be 
L_s = s! K_{(s)} = \delta^{a_1 a_2 \cdots a_s}_{b_1 b_2 
\cdots b_s} K^{b_1}{}_{a_1} K^{b_2}{}_{a_2} \cdots
K^{b_s}{}_{a_s}. 
\label{lbi}
\ee
These symmetric products of the extrinsic curvature are 
nothing but the fundamental invariants of the $(p+1)$-dimensional 
world volume $m$, referred to as the \textit{Lovelock 
type brane invariants} (LBI),~\cite{Rojas2013}. The 
first LBI are given by
\beq
L_0 &=& 1,
\label{Ls0}
\\
L_1 &=& K,
\label{Ls1}
\\
L_2 &=& K^2 - \SS = \mathcal{R},
\label{Ls2}
\\
L_3 &=& K^3 - 3K \SS + 2 K^{ab} \SS_{ab},
\label{Ls3}
\\
L_4 &=& K^4 - 6 K^2 \SS + 8 K K^{ab} \SS_{ab}
+ 3 \SS^2 - 6 \SS_{ab} \SS^{ab},
\n
\\
&=& \mathcal{R}^2 - 4 \mathcal{R}^{ab} \mathcal{R}_{ab}
+ \mathcal{R}_{abcd} \mathcal{R}^{abcd}, 
\label{Ls4}
\\
L_5 &=& K^5 - 10 K^3 \SS + 20 K^2 K^{ab} \SS_{ab}
- 30 K \SS^{ab} \SS_{ab} 
\n
\\
&+& 15 K \SS^2 - 20 \SS K^{ab} \SS_{ab} + 24 K^a{}_b 
\SS^b{}_c \SS^c{}_a,
\label{Ls5}
\eeq
where $\R$ stands for the world volume Ricci 
scalar defined on $m$ and $\SS := g^{ab} \SS_{ab}$.
In arriving at these geometric structures we have 
repeatedly used the Gauss-Codazzi integrability conditions 
in a flat background spacetime, $\R_{abcd} = K_{ac}K_{bd}
- K_{ad} K_{bc}$~\cite{Spivak1970,defo1995}.

Sticking to matrix techniques, we introduce the important 
tensors in connection to the cofactors associated to the 
scalars~(\ref{lbi}). We find that
\be 
\label{Js}
J^a_{(s)\,b} := \delta^{a a_1 a_2 \cdots a_s}_{b b_1 b_2
\cdots b_s} K^{b_1}{}_{a_1} K^{b_2}{}_{a_2} \cdots
K^{b_s}{}_{a_s},
\ee
are symmetric and divergence-free since $\nabla_a J^{ab}_{(s)}
= 0$ holds when the ambient spacetime is Minkowski~\cite{Rojas2013}. 
These are referred to as \textit{Lovelock type brane tensors}
(LBT), and they represent an extension to extended objects of 
arbitrary dimensions of the original Lovelock 
tensors~\cite{Lovelock1971}. It is worthwhile to mention that tensors~(\ref{Js}) satisfy the mighty identity 
$J^{ab}_{(s)} = g^{ab} L_s - s K^a{}_c J^{bc}_{(s)}$. 
This relationship can be proved by expanding 
out the cofactor indicated in~(\ref{Js}) in terms of minors 
as follows
\beq 
J^a_{(s)\,b} &=& [\delta^a{}_b \delta^{a_1 a_2 \cdots a_s}_{b_1 b_2
\cdots b_s} - \delta^a{}_{b_1} \delta^{a_1 a_2 \cdots a_s}_{b b_2
\cdots b_s} + \cdots 
\n
\\
&+& (-1)^s \delta^a{}_{b_s}
\delta^{a_1 a_2 \cdots a_s}_{bb_1 \cdots b_{s-1}}]K^{b_1}{}_{a_1}
K^{b_2}{}_{a_2} \cdots K^{b_s}{}_{a_s},
\n
\\
&=& \delta^a{}_b ( \delta^{a_1 a_2 \cdots a_s}_{b_1 b_2
\cdots b_s} K^{b_1}{}_{a_1}
K^{b_2}{}_{a_2} \cdots K^{b_s}{}_{a_s}) 
\n
\\
&-&  s \delta^a{}_{b_1} K^{b_1}{}_{a_1} (\delta^{a_1 a_2 
\cdots a_s}_{b b_2 \cdots b_s} K^{b_2}{}_{a_2} \cdots 
K^{b_s}{}_{a_s} ),
\eeq
where we used the index-renaming and skew-symmetric properties
of the gKd, and the definitions~(\ref{lbi}) and~(\ref{Js}) with appropriate values for $s$.

The first LBT are given explicitly by
\beq 
J^{ab}_{(0)} &=& g^{ab} = -2 G^{ab}_{(0)},
\label{Js0}
\\
J^{ab}_{(1)} &=& g^{ab} L_1 - K^{ab},
\label{Js1}
\\
J^{ab}_{(2)} &=&  -2 G^{ab}_{(1)},
\\
J^{ab}_{(3)} &=& g^{ab} L_3 - 3 \R K^{ab}
+ 6 K \mathsf{S}^{ab} - 6 K^a{}_c \mathsf{S}^{bc},
\label{Js2}
\\
J^{ab}_{(4)} &=& - 2 G^{ab}_{(2)},
\eeq
where $G^{ab}_{(n)}$ stands for the original form of 
the Lovelock tensors in pure gravity~\cite{Lovelock1971,Rojas2013}, 
where Einstein tensor $G^{ab}_{(1)}$ is included. 
This compact notation is useful for writing large explicit 
expressions since, for example, $-2G^{ab}_{(2)} = g^{ab}L_4 - 
4 (\R \R^{ab} - 2 \R^a{}_c \R^{bc} - 2 \R^{acbd} \R_{cd} + 
\R^{acde}\R^b{}_{cde})$.
On contracting these with the extrinsic curvature yields
\be 
J^{ab}_{(s)} K_{ab} = L_{s+1}.
\label{idL}
\ee
We conclude this brief survey on the basics of LBG by 
pointing out that the action functional of the 
$(p+1)$th LBI,
\be 
S[X^\mu] = \int_m \sqrt{-g} \,\det (K^a{}_b),
\label{GBterm}
\ee
is nothing but the Gauss-Bonnet topological invariant
which, as discussed in~\cite{Capovilla1996}, corresponds
to a conformal invariant functional with respect to
conformal transformations of the geometry of $m$.
For example, for $p=3$ we have that $\det (K^a{}_b) = 
\R^2 - 4 \R_{ab} \R^{ab} + \R_{abcd} \R^{abcd}$, which 
does not contribute to the corresponding equation of 
motion.

Regarding the variation of the action~(\ref{action1}),
as discussed in~\cite{Rojas2013,Rojas2016}, the Lovelock type 
brane equation results in a weaker equation in comparison 
with the original Lovelock equations, and take the compact 
form 
\be 
\label{eom1}
\sum_{s=0}^p \frac{\alpha^s}{s!} J^{ab}_{(s)} 
K_{ab} = \sum_{s=0}^p \frac{\alpha^s}{s!}L_{s+1}
= 0,
\ee
where we have used~(\ref{idL}). Although the 
action~(\ref{action1}) is of second order in derivatives 
of $X^\mu$, equation~(\ref{eom1}) represents a 
second-order in derivatives equation of motion, which 
is a signal that we have only one degree of freedom as 
result of the geometrical transverse motion. 
An apparent mistake is encountered in the upper limit 
of the series~(\ref{eom1}) in comparison with the eom reported 
in~\cite{Rojas2013,Rojas2016}, but there is no oversight 
since for such a value the corresponding LBI vanishes 
identically.

It follows from~(\ref{idL}) that the equation 
of motion~(\ref{eom1}) for $m^*$ becomes
\be 
K + \alpha \R + \frac{\alpha^2}{2} L_3 + \cdots 
+ \frac{\alpha^p}{p!} L_{p+1}  = 0.
\ee
This must be interpreted as follows. The world volume that
extremizes~(\ref{action0}), and~(\ref{action1}), in terms 
of the field variables $X^\mu$, is a minimal timelike 
hypersurface written in terms of elementary polynomials 
given by appropriate products of the fundamental forms~(\ref{FSff}) and~(\ref{Tff}). This represents a generalization of the 
well-known condition for extremal hypersurfaces in the 
sense that the vanishing of the trace of the extrinsic 
curvature is corrected by a finite series of geometric 
polynomials leading to a second-order equation of motion. 
On this basis, the case of $p=3$ is of a particular
interest. In such a case the eom reads
\be 
\begin{aligned}
K &+ \alpha \R + \frac{\alpha^2}{2} \left( K^3 
- 3K \SS + 2 K^{ab} \SS_{ab} \right)
\\
&+ \frac{\alpha^3}{6} \left( \R^2 - 4 \R_{ab}\R^{ab} 
+ \R_{abcd} \R^{abcd} \right) = 0.
\end{aligned}
\ee
Obviously, this equation might be reduced in complexity 
if $\alpha$ represents a small scale; in such a case the 
last terms would not contribute significantly.

\subsection{Connection between LBG and a DNG theory}

We now turn to a closer look at the connection between the 
LBG and the DNG theory provided by~(\ref{action0}).
Firstly, we need to compute explicitly the inverse
of the matrix $\Lambda^a{}_b$. Continuing with the use of 
the gKd symbol methods to calculate the inverse matrix 
associated to a given one~\cite{Grinfeld2013}, 
we have that
\be 
\bL^a{}_b =\frac{1}{\Lambda} \sum_{s=0}^{p} 
\frac{\alpha^s}{s!} J^a_{(s)b},
\label{invLambda}
\ee
where $\Lambda$ stands for $\det (\Lambda^a{}_b)$, and 
we have considered the tensors~(\ref{Js}). In 
passing, by applying again the gKd techniques, we can 
deduce an expression for $\Lambda$. We find
\be
\Lambda = \det (\Lambda^a{}_b) = \sum_{s=0}^{p+1} 
\frac{\alpha^s}{s!} L_s.
\label{Ldet}
\ee

\sk
Now, on according to~(\ref{ids1}), we can determine
the inverse of the starred metric $g^*_{ab}$,
\be 
\bg^{*ab} = \frac{1}{\Lambda^2} \sum_{r=0}^p \sum_{s=0}^p
\frac{\alpha^r}{r!} \frac{\alpha^s}{s!} J^a_{(r)c} J^{bc}_{(s)}.
\label{inv-g1}
\ee
We find it convenient to rewrite this expression in a 
more tractable form. Starting with the contraction 
$\frac{1}{\Lambda} \left[ \sum \frac{\alpha^n}{n!} (n+1) 
J^{ac}_{(n)}\right]g^*_{cb}$ and 
using~(\ref{ids1}), after a lengthy but straightforward computation, expression~(\ref{inv-g1}) takes a form that 
depends linearly on the conserved tensors,
\be 
\begin{aligned}
\bg^{*\,ab} &= \frac{1}{\Lambda} \sum_{r=0}^p \frac{(r+1)
\alpha^r}{r!} J^{ab}_{(r)} 
\\
&- \frac{1}{\Lambda^2} \left(\sum_{r=0}^p
\frac{\alpha^{r+1}}{r!} L_{r+1} \right) \left(
\sum_{s=0}^p \frac{\alpha^s}{s!} J^{ab}_{(s)} \right).
\end{aligned}
\label{inv-2}
\ee

The equation of motion viewed from the parallel 
world volume $m^*$, taken together with~(\ref{Kabstar}) 
and~(\ref{ids1}), yields
\be
\bg^{*ab} K^*_{ab} = \bg^a{}_c \bg^b{}_d g^{cd} \Lambda^e{}_a
K_{eb} = \bL^a{}_c g^{bc} K_{ab}.
\n
\ee
We substitute now expression for the inverse 
$\bL^a{}_b$,~(\ref{invLambda}), to obtain
\be 
\bg^{*ab} K^*_{ab} = \frac{1}{\Lambda} \sum_{s=0}^p 
\frac{\alpha^s}{s!} J^{ab}_{(s)} K_{ab}.
\n
\ee
Therefore, 
\be 
\bg^{*ab} K^*_{ab}  = 0 
\quad \Longrightarrow \quad 
\sum_{s=0}^p 
\frac{\alpha^s}{s!} J^{ab}_{(s)}K_{ab} = 0,
\label{eom3}
\ee
as expected.

Regarding the conserved stress tensor~(\ref{f-0}), when 
trying to view it from the framework of the seminal 
world volume $m$, $f^{*\,a\mu} \longrightarrow f^{a\,\mu}$, 
it can be computed directly as follows. On recalling 
again the identities~(\ref{Emua}) and~(\ref{ids1}), we have
\beq 
f^{a\,\mu} &=& - \mu \sqrt{-g^*} \,\bL^a{}_c \bL^b{}_d g^{cd}
\,\Lambda^e{}_b e^\mu{}_e,
\n
\\ 
&=& - \mu \sqrt{-g} \Lambda\,\bL^a{}_c g^{cb} \,e^\mu{}_b,
\n
\eeq
where we have used the property discussed 
below~(\ref{ids1}) for $g^*$, as well as the fact 
that $\bL^a{}_c \Lambda^c{}_b = \delta^a{}_b$.
Thus, by inserting~(\ref{invLambda}) into the
latter expression we find
\be 
f^{a\,\mu} = - \mu \sqrt{-g} \sum_{s=0}^p 
\frac{\alpha^s}{s!} J^{ab}_{(s)}\,e^\mu{}_b,
\label{fconserved2}
\ee
where we have accommodated the tensors $J^{ab}_{(n)}$ on
the dynamics of $\Sigma$. This relationship can be 
straightforwardly obtained using the approach developed
in~\cite{Capo2000} by identifying the Lagrangian function
involved in~(\ref{action1}). Notice that this conserved
stress tensor is purely tangential to $m$, as expected.
From this, trough the divergence-free property of the
tensors $J^{ab}_{(n)}$, we immediately infer the equation 
of motion~(\ref{eom1}) as well as the geometrical fact 
that $\na J^{ab}_{(s)} = 0$, as a consequence of the reparametrization invariance of the manifold $m$.

One can go one step further and obtain the linearization 
of the eom for the LBG from the previous results. Recall 
first that for a minimal hypersurface $m^*$, the equation 
of motion $K^* = 0$ gets linearisation, about a solution 
of the equation of motion, as~\cite{defo1995}
\be 
g^{*\,ab} \left( \nabla^*_a \nabla^*_b \phi 
+ \SS^*_{ab}\, \phi \right) = 0.
\label{leom-1}
\ee 
On expressing this in terms of the 
relationships~(\ref{Sabstar}) and~(\ref{Gamma2a}) 
we obtain
\beq 
0&=& \bL^a{}_c \bL^b{}_d g^{cd} \left[
\partial_a \partial_b \phi - \left( \Lambda^e{}_b \bL^f{}_h 
\Gamma^h_{ae} + \bL^f{}_h \partial_a \Lambda^h{}_b 
\right)\partial_e \phi 
\right. 
\n
\\
&& \left. \qquad \qquad \quad +\, \SS_{ab} \,\phi 
\right],
\n
\\
&=& \bL^a{}_c g^{cd} \left[ 
\bL^b{}_d \partial_a \partial_b \phi - \bL^e{}_f \Gamma^f_{ad}
\partial_e \phi + \partial_a \bL^e{}_d \,\partial_e \phi
\right.
\n
\\
&& \left. \qquad \qquad \quad +\, \bL^b{}_d 
\SS_{ab} \,\phi \right],
\n
\eeq
where we have used the relation $\bL^a{}_c \Lambda^c{}_b 
= \delta^a{}_b$ as well as relabeling some indices. 
Rewriting and simplifying the terms leads us to
\beq 
0 &=& \bL^a{}_c g^{cd} \left[ \partial_a \left( \bL^b{}_d
\partial_b \phi \right) - \Gamma^f_{ad} \bL^e{}_f\,
\partial_e \phi + \bL^b{}_d \SS_{ab}\,\phi \right],
\n
\\
&=& \bL^a{}_c g^{cd} \na \left( \bL^b{}_d \nb \phi
\right) + \bg^{*\,ab} \SS_{ab}\,\phi,
\n
\eeq
where we have used relation~(\ref{ids1}) again. 
Expression~(\ref{invLambda}) for $\bL^a{}_b$ 
allows us to rearrange the latter equation as
\be 
0 = \frac{1}{\Lambda} \left[ \na \left( \Lambda \bg^{*\,ab}
\,\nb\, \phi \right) \right] +  \bg^{*\,ab} \SS_{ab} \,\phi,
\n
\ee
where once again we considered~(\ref{ids1}). We 
can write this equation, in accordance with the 
eom~(\ref{eom1}), as follows
\be 
\frac{1}{\Lambda} \na \left[ \left( \sum_{s=0}^p 
\frac{\alpha^s}{s!} (s+1) J^{ab}_{(s)}\right) 
\nb \phi \right] + \bg^{*\,ab}\SS_{ab} \,\phi 
= 0.
\n
\ee 
By appealing again the conservation of the 
$J^{ab}_{(n)}$, we get
\be 
\frac{1}{\Lambda} \sum_{s=0}^p 
\frac{\alpha^s}{s!} (s+1) J^{ab}_{(s)}\,\na \nb \phi
+ \bg^{*\,ab}\SS_{ab} \,\phi 
= 0.
\n
\ee
That is, $\bg^{*\,ab} \left( \na \nb \phi + \SS_{ab} 
\,\phi \right) = 0$. Finally, we see that this 
relationship takes the compact form
\be 
\sum_{s=0}^p \frac{\alpha^s}{s!} (s+1) \left[ J^{ab}_{(s)}
\na \nb \phi + M^2_{(s)}\,\phi \right] = 0,
\label{Jacobi}
\ee
in complete agreement with the Jacobi equation derived 
in~\cite{Rojas2016}, which is considered as the equation 
of motion for the small perturbations $\phi$. 
The quantity
\be 
M^2_{(s)} := J^{ab}_{(s)} \SS_{ab} = \frac{1}{s+1} 
\left( L_1 L_{s+1} - L_{s+2}\right),
\label{mass}
\ee
written in terms of the LBI, plays the role of a geometric 
mass-like term.

To close this section we provide below a diagram showing the 
interplay between the two geometric points of view that 
are linked by virtue of~(\ref{sembedding}) and anchored by the 
dependence on the same coordinates, in adherence to the 
parallel surfaces framework.
It is observed that the variational processes with respect
to the embedding functions $X^{*\mu}$, followed by the change
$X^{*\mu} \rightarrow X^\mu$ commute, that is, both paths lead to 
the same outcome
\begin{figure}[ht]
    \centering
    \includegraphics[width=0.6\columnwidth]{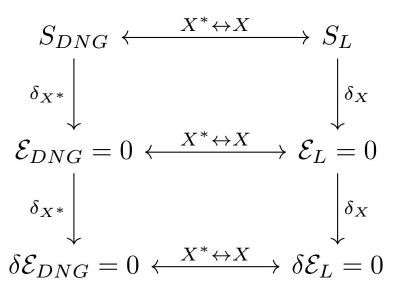}
    \label{fig:conmutativo}
\end{figure}
Here, $ S_{\text{\tiny DNG}}$,  $\mathcal{E}_{\text{\tiny DNG}}$, 
and $\delta \mathcal{E}_{\text{\tiny DNG}}$ represent 
the action, equation of motion, and the linearized 
equation, respectively, of the DNG model. Similarly, 
$S_{\text{\tiny L}}$, $\mathcal{E}_{\text{\tiny L}}$, 
and $\delta \mathcal{E}_{\text{\tiny L}}$ correspond to 
the action, equation of motion, and the linearized equation,
respectively, for a Lovelock-type brane. Finally, $\delta_{X^*}$
and $\delta_X$ denote the variations with respect to $X^{*\mu}$ 
and $X^\mu$, respectively.

\section{Inclusion of matter}
\label{sec5}

If an action matter is included in our description, 
$S_{\text{\tiny m}} = \int_m \sqrt{-g} L_{\text{\tiny m}}$, 
with a matter Lagrangian $L_{\text{\tiny m}} (\varphi (x^a), 
X^\mu)$ where $\varphi (x^a)$ denotes matter fields living 
on the brane, the form of the eom~(\ref{eom1}) remains 
practically unchanged since it only receives an extra 
contribution. Certainly, a variational process applied 
to $S_{\textit{\tiny m}}$ yields $\delta S_{\textit{\tiny m}} 
= \int_m \left[ \partial (\sqrt{-g} L_{\text{\tiny m}})/ 
\partial g^{ab} \right] \delta g^{ab}$. 
After adding this to the variation of the original 
Lovelock type brane gravity action followed by insertion 
of~(\ref{var1d}), as well as neglecting a surface boundary 
term, we find the equation of motion
\be 
\left( \sum_{s=0}^p \frac{\alpha^s}{s!} J^{ab}_{(s)}
 - T^{ab}_{\text{\tiny m}} \right) K_{ab} = 0,
\label{eom-matt}
\ee
where $T_{ab}^{\text{\tiny m}} = - (2/\sqrt{-g}) \partial
(\sqrt{-g} L_{\text{\tiny m}})/\partial g^{ab}$ is the
world volume energy-momentum tensor.

Apparently we can ensure that this theory possesses a 
built-in Lovelock limit since every solution of pure Lovelock 
equations, $G^L_{ab} - T^{\text{\tiny m}}_{ab} = 0$, is 
necessarily a solution of the Lovelock type brane gravity, 
but this observation is deceptive because in our framework 
we have a double number of conserved tensors contrary to 
what occurs in pure Lovelock theory. Despite of this, we 
may speculate on some cosmological implications in brane 
world scenarios that may arise in LBG. Guided by Davidson 
proposal about the existence of exotic matter, different from 
that coming from $L_{\text{\tiny m}}$~\cite{Davidson2003}, 
we could conjecture that~(\ref{eom-matt}) is weaker in the 
sense that a more general solution of the form $\sum_{s=0}^p 
\frac{\alpha^s}{s!} J^{ab}_{(s)} - T^{ab}_{\text{\tiny m}}  =: 
\mathcal{T}^{ab}$ may exist as long as
\be 
\mathcal{T}_{ab} K^{ab} = 0
\quad \text{and} \quad 
\mathcal{T}_{ab} \neq 0. 
\ee
As discussed in~\cite{Davidson2003,Davidson2006,Paston2018} for 
geodetic brane gravity, in our approach $\mathcal{T}_{ab}$ is 
also susceptible to being interpreted as a non-ordinary matter 
contribution, also labeled as \textit{dark matter}, or 
\textit{embedding matter}, since it is not included in the 
standard matter contribution $T_{ab}^{\text{\tiny m}}$.

In the braneworld scenarios, we further observe that for $p=3$, 
and $N=5$, taking into account the local isometric embedding 
theorem~\cite{Friedman1961,Rosen1965}, we have an effective 
action yielding second-order equation of motion, which 
is also susceptible to being described as a DNG type action 
in a parallel world volume $m^*$ laid off equal distance 
$\alpha$ along the normal $n^\mu$ associated to $m$
\be
S[X^\mu] = \alpha_0 \int_m \sqrt{-g} \sum_{n=0}^3 
\frac{\alpha^n}{n!} L_n,
\label{action3}
\ee
where $\alpha_0$ is a constant with appropriate dimensions. 
Explicitly,
\beq 
S[X^\mu] &=& \int_m \sqrt{-g} \left[ \alpha_0 + \beta\,K +
\kappa\,\R 
\right.
\n
\\
& & \quad \qquad + \left. \gamma \left( K^3 - 3K\,\SS 
+ 2 K^{ab} \SS_{ab} \right) \right],
\label{action4}
\eeq
where $\beta:= \alpha_0 \alpha$, $\kappa:= \alpha_0 
\alpha^2/2$ and $\gamma := \alpha_0 \alpha^3/6$. The first 
three terms have been studied in some contributions~\cite{Rojas2012,Rojas2014,Rojas2024}, but 
considering independent values for the parameters 
accompanying each LBI providing a peculiar acceleration 
behaviour, and reproducing cosmological effects arising 
in other modified gravity theories. We now wonder if the 
last term could reproduce or mimic some effects arising  
from other alternative modified theories. In this sense, 
our guess is that the fundamental invariant $L_3$ will 
reproduce acceleration effects of Gauss-Bonnet cosmology~\cite{Charmousis2002,Bouhmadi2008}. 
On the contrary, if we are interested in quantum 
approximations for LBG, it could be convenient to set our 
theory in $m^*$ and then analyze a DNG type action to 
subsequently apply known quantization techniques and then 
explore the quantum correspondence between the theories. 
All this deserves further study which will be reported 
elsewhere.

\subsection{Varying distance for neighbouring world volumes}

Another approach worth exploring is the extension of 
our analysis to the description of neighbouring 
world volumes with a varying distance along the 
normal $n^\mu$, that is, $\alpha \longrightarrow 
\Phi (X^{\mu}(x^{a}))=\Phi(x^a)$. The world volume $m^*$ can be 
parameterized as follows
\be 
X^{*\,\mu} (x^a) = X^\mu (x^a) + \Phi (x^a)\,n^\mu (x^a).
\label{Xvarphi}
\ee 
The tangent space of $m^*$ is spanned now by the vectors
\be 
E^\mu{}_a = \Lambda^b{}_a\,e^\mu{}_b + \Phi_a \,
n^\mu,
\label{Emua2}
\ee 
where $\Phi_a := \partial_a \Phi$, and $\Lambda^a{}_b$
is defined in~(\ref{matrixL}). The corresponding induced 
metric now reads
\beq 
g^*_{ab} &=& \left( \Lambda^c{}_a \,e_c + \Phi_a \,n
\right)\cdot \left( \Lambda^d{}_b \,e_d + \Phi_b \,n\right),
\n
\\
&=& \Lambda^c{}_a \Lambda^d{}_b g_{cd} 
+ \delta^c{}_a \delta^d{}_b \,\Phi_c \Phi_d.
\label{disformal}
\eeq
We intend now the inverse matrix $\bL^a{}_b$ to enter the 
game as follows. By suitably writing the latter equation
we get
\beq
g^*_{ab} &=& \Lambda^c{}_a \Lambda^d{}_b \left( g_{cd} +
\bL^e{}_c \bL^f{}_d\,\Phi_e \Phi_f  \right),
\n
\\
&=& \Lambda^c{}_a \Lambda^d{}_b\,g_{ch} \left( 
\delta^h{}_d + g^{hl} \bL^e{}_l \bL^f{}_d\,\Phi_e 
\Phi_f  \right).
\n
\eeq
It is straightforward to rearrange this relationship 
as
\be 
g^*_{ab} = g_{cf} \Lambda^c{}_a \Lambda^d{}_b \left(
\delta^f{}_d + \mathcal{K}^f{}_d \right),
\label{gstar2} 
\ee
where we have introduced the cumbersome matrix
\beq 
\mathcal{K}^a{}_b &:=& g^{ac} \bL^e{}_c \bL^f{}_b \,
\Phi_e \Phi_f,
\n
\\ 
&=& \frac{1}{\Lambda^2} \sum_{r=0}^p \frac{\Phi^r}{r!}
J^{ac}_{(r)} \sum_{s=0}^p \frac{\Phi^s}{s!} J^d_{(s)b}\,\,
\Phi_c \Phi_d.
\label{gstar3}
\eeq
Here, $\Phi^r$ and $\Phi^s$ denote the scalar field powered 
in $r$ and $s$, respectively. Hence, from the well known 
identity for the product of determinants we have
\be 
\sqrt{-g^*} = \sqrt{-g}\,\Lambda \sqrt{\det \left( \delta^a{}_b 
+ \mathcal{K}^a{}_b \right)}.
\label{gstar4}
\ee
As for the remaining fundamental forms, they do not undergo 
any change due to the choice of~(\ref{Xvarphi}). 
From~(\ref{gstar3}) is quite obvious the high degree of 
complexity when establishing a starred DNG and desire to cast 
out interesting physical implications. Nevertheless, from our 
previous results and concerning the framework of the so-called 
ruled surfaces, we can make contact with specializations 
leading to physical implications. Indeed, for $p=2$ at the 
Euclidean context, if $\varphi (x^a) = x^2$, and $X^{*\mu}
(x^0,x^1) = X^\mu(x^0,x^1) + x^2 n^\mu (x^0,x^1)$ our previous 
expression is specialized to the treatment to obtain either a 
Schr\"{o}dinger equation or a Dirac 
equation~\cite{Ferrari2008,Brandt2016} on curved surfaces 
describing a particle subject to physical fields.
Some remarks are in order. Expression~(\ref{disformal}) 
defines a particular form of a disformal transformation 
between the geometries of $m$ and $m^*$ via $g_{ab}$, 
$g^*_{ab}$ and the scalar field $\Phi$, which characterizes 
the presence of scalar-tensor theories closely related to 
Horndeski or Galileon
theories~\cite{Bekenstein1993,Zumala2014,Rojas2015,Deffayet2020}, 
and which might offer insights into the peculiar dark 
energy/matter type contributions arising in brane world 
scenarios. Second, $\Phi \longrightarrow \alpha$ marks 
the parallel world volume limit. Indeed, such a limit 
takes us back to the expression outlined below~(\ref{ids1}).

\section{Conclusions}
\label{sec6}

In this paper we have formally derived the Lovelock type 
brane gravity from a DNG action within the parallel surfaces 
framework adapted to extended objects. Our approach justifies 
and establishes, through geometrical and matrix techniques, 
the underpinnings of the LBG as well as a correspondence 
with an extended DNG action whose volume element is built 
with the metric associated to a parallel world volume to a 
pioneer one. The actions $S[X^{*\mu}]$ and $S[X^\mu]$ characterize 
the same physical system but from different points of view. 
Indeed, both actions lead to a single second-order equation 
of motion expressed in terms of different geometries so as to 
they describe the evolution of the same degree of freedom.
We have highlighted the dependence of the intrinsic and 
extrinsic geometries of $m^*$ on the fundamental 
forms and the divergence-free tensors $J^{ab}_{(n)}$ 
associated with the geometry of the primordial manifold 
which is suitably achieved by the transformation matrix 
$\Lambda^a{}_b$~(\ref{matrixL}). The inclusion of matter 
is direct without markedly affecting mathematical development.
In the framework of parallel world volumes discussed in 
this work, notice that there is not an endless number of parallel 
world volumes to a given one since this family 
depends of the dimension of the primordial world volume, 
so there is a series expansion that must be finite. Indeed, 
for instance, on cosmological context this last point is 
important since the series expansion in (\ref{action1}), 
arising from the volume element in~(\ref{action0}), is 
finite. The parallel surface framework has cropped up in 
other interesting contexts. Certainly, in a Euclidean scenario, 
specifically in the framework of ruled surfaces which is 
related to the parallel surfaces framework, by choosing a 
privileged direction this scheme helps to obtain a 
Schr\"{o}dinger equation or a Dirac equation on a curved surface~\cite{Ferrari2008,Brandt2016}. In a like manner, 
in our brane gravity theory, by extending the ruled surfaces 
approach for extended objects, we have the presence of two 
geometries in a single brane theory which manifests itself 
in the appearance of a disformal transformation. We believe 
that this gives a clue to the existence of some type of 
Galileons, which will be explored elsewhere.

As in the case of geodetic brane gravity, LBG modifies 
pure Lovelock gravity, and allows for the appearance of 
additional energy/matter in contrast with ordinary matter, 
making it an alternative to explain the dark matter observed 
in pure gravity on merely geometrical grounds. 
Next task is to explore the dark energy 
content within the LBG framework according to 
$\mathcal{T}^{ab}$ by considering one single parameter 
$\alpha$ contrary to the approaches adopted in~\cite{Rojas2012,Rojas2014,Rojas2024}. 
What is remarkable to observe is how the correction terms 
associated with the extrinsic curvature of the brane,  
included in the expansion~(\ref{action1}), can reproduce 
many of the general features of the late acceleration 
behaviour for our universe.
Our approach also aims to exploit the correspondence 
between LBG and a DNG theory to make contact with already 
known Hamiltonian approaches for a DNG setup in order to 
advance the exploration of Hamiltonian approximations for LBG, 
thus avoiding the use of a tedious Ostrogradski-Hamilton 
framework. All of the above is in the interest of entering into 
quantum approaches that can be applied primarily to brane 
cosmology. This will be reported in forthcoming works.  

\begin{acknowledgments}
ER is grateful to R. Cordero for valuable comments. 
ER acknowledges encouragement from 
ProDeP-M\'exico, CA-UV-320: \'Algebra, Geometr\'\i a y 
Gravitaci\'on. GC acknowledges support from a Postdoctoral 
Fellowship by Estancias Posdoctorales por M\'exico 
2023(1)-CONAHCYT . Also, ER thanks partial support 
from Sistema Nacional de Investigadoras e Investigadores, 
M\'exico.
\end{acknowledgments}

\appendix

\section{On the equation of motion $K^* = 0$}
\label{app1}

The variation of~(\ref{action0}) with respect to the field 
variables $X^\mu$, on recalling the 
usual identity $\delta (\sqrt{-g}) = (1/2) \sqrt{-g} 
g^{ab}\delta g_{ba}$, becomes
\beq 
\delta S 
&=& - \mu \int_{m^*} \frac{1}{2} \sqrt{-g^*}
\bg^{*ab} \,\delta g^*_{ab},
\n
\\
&=& - \mu \int_{m^*} \sqrt{-g^*}
\bg^{*ab}  \left( K^*_{ab}\,\phi 
- \alpha\, \Lambda^c{}_{a} \nabla_{b} \nc \phi \right)
\n
\\
&-& \frac{\mu}{2} \int_{m^*} \sqrt{-g^*} \bg^{*ab}
\Ld g^*_{ab},
\n
\eeq
where we have used~(\ref{var2a}). Regarding the 
second integral, named $\delta_\parallel S$, on recalling 
that the Lie derivative operation is connection independent, 
and from the fundamental theorem of Riemannian geometry, 
namely, if $g^*_{ab}$ is a metric tensor, there exists a 
unique symmetric connection $\nabla^*_a$ such that 
$\nabla^*_a g^*_{bc} = 0$, then
\beq 
\delta_\parallel S &:=& - \mu  \int_{m^*} \sqrt{-g^*} 
\bg^{*ab} \,\nabla^*_a \phi_b,
\n
\\
&=&  \int_{m^*} \partial_a \left( - \mu \sqrt{-g^*} 
\bg^{*ab} \phi_b \right).
\n
\eeq
That is, the tangential variation provides a merely
boundary term which does not contribute to the eom as 
a consequence of the invariance under reparametrizations 
of $m^*$.

Let us now focus on the contribution to the variation of the 
second term in the first integral. By expanding the covariant 
derivative followed of integrating by parts, as well as 
collecting all terms and relabelling the indices, we get
\beq 
\delta S_1 
&=& \int_{m^*} \partial_a \left[
\partial_b \left( \sqrt{-g^*} \bg^{*bc} \right)\Lambda^a{}_c
+ \sqrt{-g^*} \bg^{*bc}\,\Lambda^d{}_b \Gamma^a_{cd}
\right]\phi
\n
\\
&+& \int_{m^*} \partial_a \widetilde{T}^a,
\n
\eeq
with $\widetilde{T}^a := \left( \sqrt{-g^*} \bg^{*bc} 
\Lambda^a{}_b \right)\partial_c\phi - \partial_b \left( 
\sqrt{-g^*} \bg^{*ac} \Lambda^b{}_c \right) \phi - \sqrt{-g^*} 
\bg^{*bc} \Lambda^d{}_b \Gamma^a_{cd} \,\phi $. Up to 
a total derivative, considering the identity $\partial_b 
\left( \sqrt{- g^*} \bg^{*ab}\right) = - \sqrt{-g^*} 
\bg^{* bc} \Gamma^{*a}_{bc}$, and a full arrangement of 
the various terms, one can readily check that $\delta S_1$ 
reduces to
\be
\delta S_1 = \int_{m^*} \partial_a \left[ \sqrt{-g^*} \bg^{*bc} 
\left( -  \Lambda^a{}_d \Gamma^{*d}_{bc} + \Lambda^d{}_b
\Gamma^a_{cd} +  \partial_b \Lambda^a{}_c \right) 
\right]\phi
\n
\ee
which vanishes identically after substitution of 
expression~(\ref{Gamma2a}) defining the starred connection. 

We have therefore that the variation of action~(\ref{action0})
leads to 
\be 
\delta S = - \mu \int_{m^*} \sqrt{-g^*} \bg^{*ab}K^*_{ab}\,\phi.
\ee
Therefore, as a classical equation of motion, we obtain a minimal 
surface condition for $m^*$ in terms of its geometry
\be 
K^* := \bg^{*ab}K^*_{ab} = 0,
\ee
as expected. 

\section{Connecting up $\sqrt{- g^*}$ and $\sqrt{-g}$} 
\label{app2} 

Here we outline the derivation of~(\ref{action1}) from~(\ref{action0}).
The starting point in the proof relies in the definition of the
determinant of a $(n\times n)$ matrix, $A^a{}_b$, in terms of the gKd
\be 
A:= \det\left( A^a{}_b \right) = \frac{1}{n!}\delta^{a_1a_2
\cdots a_n}_{b_1 b_2\cdots b_n} A^{a_1}{}_{b_1} 
A^{a_2}{}_{b_2} \cdots A^{a_n}{}_{b_n}.
\ee
By inserting the matrix $M^a{}_b := \delta^a{}_b + \alpha 
A^a{}_b$ in the above expression, with $\alpha$ being an 
arbitrary parameter, and $M:= \det (M^a{}_b)$, we get
\be 
n! \,M = \delta^{a_1a_2
\cdots a_n}_{b_1 b_2\cdots b_n} (\delta^{b_1}{}_{a_1} 
+ \alpha A^{b_1}{}_{a_1} ) \cdots 
( \delta^{b_n}{}_{a_n} + \alpha A^{b_n}{}_{a_n}).
\n
\ee
This estructure is closely related to the well-known characteristic determinant~ \cite{Lovelock1989}.
When performing the products, we observe that each term that 
accompanies the powers of the parameter $\alpha$ has the form
\beq 
n!\,M &=& 
\begin{pmatrix}
n
\\
0
\end{pmatrix}
\delta^{a_1 a_2 \cdots a_n}_{a_1 a_2 \cdots a_n} \alpha^0 +
\begin{pmatrix}
n
\\
1
\end{pmatrix}
\delta^{a_1 a_2 \cdots a_n}_{b_1 a_2 \cdots a_n} 
A^{b_1}{}_{a_1} \alpha + \cdots
\n
\\
&+& 
\begin{pmatrix}
n
\\
s
\end{pmatrix}
\delta^{a_1 a_2 \cdots a_s a_{s+1} \cdots a_n}_{b_1 b_2 \cdots 
b_s a_{s+1} a_n} A^{b_1}{}_{a_1} A^{b_2}{}_{a_2}
\cdots A^{b_s}{}_{a_s} \alpha^s + \cdots
\n
\\
&+&   
\begin{pmatrix}
n
\\
n
\end{pmatrix}
\delta^{a_1 a_2 \cdots a_n}_{b_1 b_2 \cdots a_n} A^{b_1}{}_{a_1} 
A^{b_2}{}_{a_2} \cdots A^{b_n}{}_{a_n} \alpha^n,
\n
\eeq
where $\begin{pmatrix}
n
\\
s
\end{pmatrix}
= \frac{n!}{(n-s)! s!}$. Bearing in mind the identity
$\delta^{a_1 a_2 \cdots a_s a_{s+1} \cdots a_r}_{b_1 b_2
\cdots b_s a_{s+1} \cdots a_r} = \frac{(n-s)!}{(n-r)!} 
\delta^{a_1 a_2 \cdots a_s}_{b_1 b_2 \cdots b_s}$ for $r\leq 
n$, and in particular the value it acquires when 
$r=n$ given by $\delta^{a_1 a_2 \cdots a_s a_{s+1} 
\cdots a_r}_{b_1 b_2 \cdots b_s a_{s+1} \cdots a_r} = (n-s)! 
\delta^{a_1 a_2 \cdots a_s}_{b_1 b_2 \cdots b_s}$, as well
as $\delta^{a_1 a_2 \cdots a_n}_{a_1 a_2
\cdots a_n} = n!$, we obtain
\beq 
n! M &=& n! \alpha^0 + \frac{n(n-1)!}{1!} 
\delta^{a_1}{}_{b_1} A^{b_1}{}_{a_1} \alpha 
\n
\\
&+& \frac{n(n-1)(n-2)!}{2!} \delta^{a_1 a_2}_{b_1b_2}
A^{b_1}{}_{a_1}A^{b_2}{}_{a_2} \alpha^2 + \cdots
\n
\\
&+&  \delta^{a_1 a_2 \cdots a_n}_{b_1 b_2 \cdots a_n} 
A^{b_1}{}_{a_1} A^{b_2}{}_{a_2} \cdots A^{b_n}{}_{a_n} 
\alpha^n.
\n
\eeq
Therefore,
\beq 
\det (\delta^a{}_b + \alpha A^a{}_b) &=&
1 + \sum_{s=1}^n \frac{\alpha^s}{s!} \delta^{a_1 a_2
\cdots a_s}_{b_1 b_2 \cdots b_s} A^{b_1}{}_{a_1}
A^{b_2}{}_{a_2} \cdots A^{b_s}{}_{a_s}
\n
\\
&=& 1 + \sum_{s=1}^n \alpha^s\,A_{(s)},
\label{B2}
\eeq
where $A_{(s)}$, denotes the $(s \times s)$ 
principal minor of $\det (A^a{}_b)$, defined as
$s!A_{(s)} = \delta^{a_1 a_2
\cdots a_s}_{b_1 b_2 \cdots b_s} A^{b_1}{}_{a_1}
A^{b_2}{}_{a_2} \cdots A^{b_s}{}_{a_s}$.

Furthermore, according to the elementary multiplication 
property of determinants, from~(\ref{ids1}), we have $g^* 
:= \det (g^*_{ab}) = g\,\Lambda^2$ where $\Lambda:= 
\det (\Lambda^a{}_b)$. Substituting the form of the 
transformation matrix, $\Lambda^a{}_b = \delta^a{}_b + 
\alpha\,K^a{}_b$,~(\ref{matrixL}), into the result~(\ref{B2}) 
we write $\Lambda$ in the form
\be 
\Lambda = \det\left( \delta^a{}_b + \alpha K^a{}_b \right)
= 1 + \sum_{s=0}^{p+1} \alpha^s\,K_{(s)},
\ee
where $K_{(s)}$ stands for the $(s\times s)$ minor of
$\det (K^a{}_b)$.

Putting all these elements together yields
\be 
\sqrt{-g^*} = \sqrt{-g} \left( 1 + \sum_{s=0}^{p+1}
\alpha^s\,K_{(s)} \right) = \sqrt{-g} \left( 1 + \sum_{s=0}^{p+1}
\frac{\alpha^s}{s!}\,L_{s}\right),
\ee
with $L_s$ defined in \eqref{lbi}.





\bibliography{love-dng-f}

\end{document}